\documentclass[]{research}

\usepackage{algpseudocode}
\usepackage[ruled,vlined]{algorithm2e}

\usepackage{amsmath}
\usepackage{amsfonts}
\usepackage{hyperref}
\usepackage{url}
\usepackage{booktabs}
\usepackage{multirow}
\usepackage{pgfplots}
\usepackage{graphicx}
\usepackage{array}
\pgfplotsset{compat=1.18}
\usepackage{arydshln}

\usepackage{xspace}

\usepackage[normalem]{ulem}
\newcolumntype{L}[1]{>{\raggedright\let\newline\\\arraybackslash\hspace{0pt}}m{#1}}
\newcolumntype{C}[1]{>{\centering\let\newline\\\arraybackslash\hspace{0pt}}m{#1}}
\newcolumntype{R}[1]{>{\raggedleft\let\newline\\\arraybackslash\hspace{0pt}}m{#1}} 
\newcommand{\xpar}[1]{\noindent\textbf{#1}\ \ }

\newcommand{\sect}[1]{\S~\ref{sect:#1}}

\newcommand{\eqn}[1]{Equation~\ref{eq:#1}}

\newcommand{\fig}[1]{Figure~\ref{fig:#1}}

\newcommand{\tbl}[1]{Table~\ref{tab:#1}}

\newcommand{\lblfig}[1]{\label{fig:#1}}
\newcommand{\lblsect}[1]{\label{sect:#1}}

\newcommand{\lbleqn}[1]{\label{eq:#1}}
\newcommand{\lbltbl}[1]{\label{tab:#1}}

\newcommand{\ignorethis}[1]{}

\makeatletter
\DeclareRobustCommand\onedot{\futurelet\@let@token\@onedot}
\def\@onedot{\ifx\@let@token.\else.\null\fi\xspace}

\def\eg{\emph{e.g}\onedot}

\makeatother

\definecolor{citecolor}{rgb}{34,139,34}
\definecolor{mydarkblue}{rgb}{0,0.08,1}
\definecolor{mydarkgreen}{rgb}{0.12,0.7,0.12}
\definecolor{mydarkred}{rgb}{0.8,0.02,0.02}
\definecolor{mydarkorange}{rgb}{0.40,0.2,0.02}
\definecolor{mypurple}{RGB}{111,0,255}
\definecolor{myred}{rgb}{1.0,0.0,0.0}
\definecolor{mygold}{rgb}{0.75,0.6,0.12}
\definecolor{mydarkgray}{rgb}{0.66,0.66,0.66}

\definecolor{darkgreen}{rgb}{0.15, 0.75, 0.15}
\definecolor{mitblue}{rgb}{0.88,0.95,0.96}
\definecolor{lightblue}{rgb}{0.90, 0.95, 0.99}

\newcommand{\FreeToken}{FreeToken\xspace}
\newcommand{\MoE}{MoE\xspace}
\newcommand{\CPU}{CPU\xspace}
\newcommand{\GPU}{GPU\xspace}
\newcommand{\PCIe}{PCIe\xspace}
\newcommand{\VRAM}{VRAM\xspace}
\newcommand{\KVCache}{KV cache\xspace}
\newcommand{\CudaGraph}{CUDA Graph\xspace}
\newcommand{\TTFT}{TTFT\xspace}

\newcommand{\FTW}{FTW\xspace}

\newcommand{\CpuExpertPool}{CPU-resident expert pool\xspace}
\newcommand{\GpuExpertCache}{GPU expert cache\xspace}
\newcommand{\ElasticGpuExpertCache}{elastic expert cache\xspace}

\newcommand{\BandwidthAdaptiveExecution}{bandwidth-adaptive execution\xspace}
\newcommand{\OnDemandCacheFill}{on-demand cache fill\xspace}

\newcommand{\TemporalExpertLocality}{temporal expert locality\xspace}
\newcommand{\RuntimeVramBudget}{runtime VRAM budget\xspace}

\newcommand{\PureCpuMoeBackend}{pure-CPU MoE backend\xspace}

\newcommand{\SemanticAwareStateCache}{semantic-aware state cache\xspace}

\newcommand{\SemanticAnchors}{semantic anchors\xspace}
\newcommand{\HybridAttention}{hybrid-attention\xspace}

\newcommand{\ExpertBanks}{expert banks\xspace}

\newcommand{\BHost}{B_{\mathrm{H}}}
\newcommand{\BPcie}{B_{\mathrm{P}}}
\newcommand{\BResidual}{B_{\mathrm{R}}}
\newcommand{\HitSet}{\mathcal{H}}
\newcommand{\MissSet}{\mathcal{M}}
\newcommand{\FillSet}{\mathcal{F}}
\newcommand{\CpuSet}{\mathcal{C}}
\newcommand{\GpuSet}{\mathcal{G}}

\makeatletter
\def\adl@drawiv#1#2#3{%
        \hskip.5\tabcolsep
        \xleaders#3{#2.5\@tempdimb #1{1}#2.5\@tempdimb}%
                #2\z@ plus1fil minus1fil\relax
        \hskip.5\tabcolsep}
\newcommand{\cdashlinelr}[1]{%
  \noalign{\vskip\aboverulesep
           \global\let\@dashdrawstore\adl@draw
           \global\let\adl@draw\adl@drawiv}
  \cdashline{#1}
  \noalign{\global\let\adl@draw\@dashdrawstore
           \vskip\belowrulesep}}
\makeatother

\title{FreeToken: Efficient Edge-Native MoE Serving with Bandwidth-Adaptive Execution}
\author[*]{Shuo Yang}
\author[*]{Xiaoze Fan}
\author[]{Melissa Pan}
\author[]{Haocheng Xi}
\author[]{Zhe Wang}
\author[]{Shanlin Sun}
\author[]{Kurt Keutzer}
\author[]{Song Han}
\author[]{Matei Zaharia}
\author[]{Chenfeng Xu\textsuperscript{\ensuremath{\dagger}}}
\author[]{Ion Stoica\textsuperscript{\ensuremath{\dagger}}}

\affiliation[]{}

\contribution[*]{Equal Contribution}
\contribution[]{\textsuperscript{\ensuremath{\dagger}}Co-Advise}

\abstract{

Frontier open-weight models are increasingly available, but serving them still largely assumes datacenter infrastructure. We present \textbf{FreeToken}, an edge-native \MoE serving system that treats a personal machine not as a small GPU, but as a unified, elastic inference platform. FreeToken co-designs the full serving stack, including model layout and loading, expert residency, CPU--GPU execution, agentic state reuse, and runtime memory management, around two realities of local AI: agent workloads continuously change their execution pattern, and edge hardware exposes heterogeneous resources whose balance differs from machine to machine. Rather than committing to a fixed offloading strategy, FreeToken continuously maps computation and model state onto the resources actually available. FreeToken supports more than 20 \MoE models and real coding and tool-using agents across hardware ranging from an 8\,GB laptop GPU to a single workstation GPU. More importantly, it changes what these machines can practically serve, from a 35B model on a laptop to a 284B model on a gaming desktop and the 753B GLM-5.2 on a single workstation GPU. FreeToken turns open weights into deployable local software, making the machines users already own a practical platform for frontier-scale intelligence.

}

\correspondence{Shuo Yang at \href{mailto:andy_yang@berkeley.edu}{andy\_yang@berkeley.edu}, Chenfeng Xu at \href{mailto:xuchenfeng@utexas.edu}{xuchenfeng@utexas.edu}}

\metadata[Code]{\url{https://github.com/FlashML-org/FreeToken}}
\metadata[Download]{\url{https://flashml.ai}}

\begin{document}

\maketitle

\section{Introduction}
\lblsect{introduction}

Recent open-weight models, such as Kimi-K3~\citep{kimik32026}, GLM-5.2~\citep{glm522026} and DeepSeek-V4-Flash-0731~\citep{deepseekv4flash2026}, are rapidly closing the capability gap with the strongest proprietary systems. Yet releasing model parameters determines only who can obtain a model, not who can afford to run it. Frontier open models still rely on scarce datacenter-class \GPU clusters that can cost millions of dollars, and although hosted APIs may be cheaper than comparable proprietary offerings, sustained use remains expensive. As agentic applications sharply increase inference demand~\citep{gpusupply2026,claudecodecosts2026}, this cost becomes particularly burdensome for individual users and small teams. Consequently, the capability gap between open and proprietary models is closing much faster than the accessibility gap between those who can obtain frontier models and those who can use them at scale.

At first glance, this accessibility gap appears to be a hardware problem. Yet more than one hundred million consumer machines already contain discrete GPUs, spanning gaming desktops, workstations, and high-performance laptops.\footnote{Steam alone reports more than 200 million monthly active users, with discrete NVIDIA GPUs present in roughly 72\% of surveyed systems~\citep{steammau2026,steamsurvey2026}.} Collectively, these machines represent an enormous pool of capable but underutilized compute. The missing piece is therefore not hardware itself, but a serving system that can treat each heterogeneous consumer machine as a unified inference platform and automatically map its GPU, CPU, memory, and interconnect resources to the strongest model configuration it can run efficiently.

\MoE architectures open a new path toward serving frontier-scale open-weight models on edge devices. An \MoE layer contains hundreds of experts while routing each token through only a small subset. DeepSeek-V4-Flash, for example, activates 6 of 256 routed experts in each of its 43 layers, so only 13B of its 284B parameters participate in any single token. At the deployed precision, this active parameter footprint fits within the 32GB memory capacity of an RTX 5090. However, sparsity reduces per-token computation without proportionally reducing the memory required for the complete expert pool. The full model may still exceed GPU memory by a wide margin, forcing inactive experts to reside in host memory or secondary storage and enter the execution path on demand. \MoE therefore creates both the opportunity and the central systems challenge for frontier inference on personal hardware: sparse activation makes the computation feasible, while the full expert pool makes efficient serving difficult.

\begin{figure}[t]
    \centering
    \includegraphics[width=\textwidth]{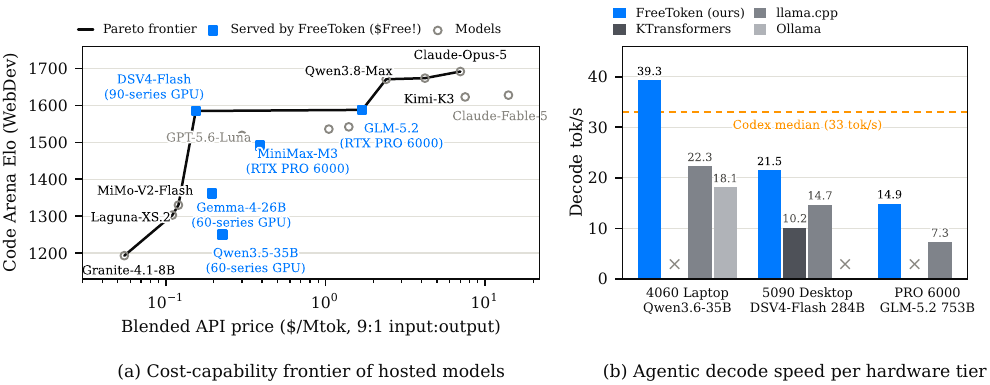}
    \caption{\FreeToken serves the models on the cost--capability
    Pareto frontier, at interactive speed on consumer hardware.
    \textbf{(a)}~Blended API list price (9:1 input:output mix,
    following the token economics measured on real coding-agent
    traces~\citep{zhu2026tracelab}) versus Code Arena
    Elo~\citep{codearena2026} for representative hosted models.
    Blue squares mark models \FreeToken serves, tagged with the
    consumer GPU class that serves them; the frontier segment from
    DeepSeek-V4-Flash to GLM-5.2 is exactly this set.  Kimi-K3
    releases open weights but exceeds consumer memory (594\,GB);
    Qwen3.5-35B stands in for its successor Qwen3.6-35B, which has
    no arena rating yet.  \textbf{(b)}~Mean decode throughput on
    real agentic workloads for the strongest model each hardware
    tier holds (coding agents on the first two tiers, a math agent
    on the third), against actively maintained edge engines.  The
    dashed line marks the median decode speed of Codex in production
    traces (33\,tok/s~\citep{zhu2026tracelab}); \(\times\) marks
    configurations an engine cannot serve.}
    \lblfig{teaser}
\end{figure}

A growing ecosystem of serving systems has begun to bring powerful open-weight models to personal hardware, including llama.cpp~\citep{gerganov2023llamacpp},KTransformers~\citep{ktransformers2025}, and Ollama~\citep{ollama2023}. Yet these existing systems address only fragments of the edge \MoE serving problem and fall short in three dimensions that prevent edge serving from reaching its theoretical capacity:
\begin{itemize}
    \item First, prefill largely destroys the working-set sparsity in \MoE. Although each token activates only a few experts, the union of routes across a long prompt often covers most experts in every layer, turning the expert working set effectively dense. This creates both computational and memory-movement challenges, as experts exceeding VRAM must be repeatedly streamed from host memory. This problem is particularly severe in agentic tool-calling workloads, where long and continuously growing contexts trigger frequent prefills; yet, existing edge serving systems provide little support for hiding expert movement or reusing recurrent state across turns.

    \item Second, decode presents the opposite regime: each token activates only a sparse subset of experts, but cache misses require experts to be repeatedly loaded, evicted, or executed from host memory. Existing systems lack a principled policy for serving these misses. Static placement cannot follow token-level routing changes, while prediction and prefetching may reduce the miss rate but do not determine how unavoidable misses should be divided between PCIe transfer, GPU execution, and direct CPU execution.

    \item Third, these two problems are amplified by the diversity and variability of edge resources. Unlike datacenter deployments, consumer hardware differs widely in GPU capacity, PCIe bandwidth, host-memory bandwidth, CPU capability, and available VRAM. These resources are also rarely dedicated exclusively to model serving: users may run browsers, games, and other applications concurrently, causing the available memory and compute budget to fluctuate over time. As a result, no single static placement or scheduling policy can perform well across devices, workload phases, and changing runtime conditions.

\end{itemize}


We propose \textbf{FreeToken}, a system built on two execution principles and one elastic resource-management policy. \textbf{(1) Bandwidth-adaptive execution} turns limited edge bandwidth from a fixed bottleneck into a runtime scheduling signal. During \textit{prefill}, FreeToken double-buffers expert movement with computation: while the \GPU evaluates the current layer, the next layer's experts stream over \PCIe. \textit{Decode} requires a more fine-grained allocation because \PCIe transfer and direct \CPU expert execution both draw from the same host-memory bandwidth. FreeToken therefore applies a $q^\star$ policy that partitions each step's cache misses between \GPU cache fill and direct \CPU execution (\sect{design-decode}), matching the bandwidth that the deployed machine can sustain. \textbf{(2) Semantic-aware caching} determines what to retain in scarce memory. Across agent turns, agentic harnesses edit context at semantic boundaries, such as thinking segments and tool calls. During \textit{prefill}, FreeToken anchors recurrent-state checkpoints at these boundaries so that, after an edit, only the new suffix is recomputed. During \textit{decode}, adjacent tokens frequently route to overlapping experts. FreeToken captures this inter-token routing locality with a shared LRU expert cache, allowing most routed accesses to hit in \VRAM and leaving only the residual misses to the $q^\star$ policy. \textbf{(3) Elastic edge resource management} adapts FreeToken to the changing memory conditions of personal hardware. At scheduler safe points, FreeToken dynamically resizes and rebuilds the \GPU expert cache under a revised memory budget without restarting the engine or reloading the host-resident expert pool. FreeToken further reduces startup latency by loading experts directly into their final host layout before pinning the populated memory.



\FreeToken supports more than 20 \MoE models across a wide range of consumer and workstation-class hardware. In the experiment section, we mainly evaluate \FreeToken on three representative frontier models, including Qwen3.6-35B-A3B, DeepSeek-V4-Flash, and GLM-5.2; six machines spanning from an 8 GB RTX 4060 laptop to a single RTX PRO 6000 workstation; using four real agentic workloads; and comparing against llama.cpp, Ollama, KTransformers, and MoE-Infinity. Specifically, on an RTX 5090, \FreeToken sustains 77--83\,tok/s on Qwen3.6-35B-A3B and 22--25\,tok/s on DeepSeek-V4-Flash, achieving 1.5--2.3x higher decode throughput than state-of-the-art edge serving across all workloads. Its performance also remains remarkably stable as workloads become increasingly agentic: the decode rate stays within 12\% of the single-turn setting, while competing systems degrade substantially. The advantage is even more pronounced in tail latency. \FreeToken keeps worst-case \TTFT below 44\,s across every workload, whereas each baseline exceeds 150\,s in at least one setting, long enough to trigger timeouts in real agent clients. Across five consumer systems, \FreeToken improves decode throughput by 1.3--2.1x. On an 8,GB RTX 4060 laptop, it serves a 35B model at 39.3\,tok/s, exceeding the 33\,tok/s median decode speed of Codex. On a 32,GB gaming desktop, it serves a 284B model interactively. On a single RTX PRO 6000 workstation GPU, it serves the 753B GLM-5.2 at twice the throughput of llama.cpp. 

Together, these advances turn open weights into open access, bringing frontier intelligence from datacenter infrastructure to the machines users already own. \FreeToken pushes both the speed and capability frontier of local inference, enabling personal hardware to interactively serve models that were previously practical only in the datacenter. We release the system at \url{flashml.ai}.

\section{Challenges in Edge MoE Serving}
\lblsect{motivation}

\MoE is a natural architecture for edge serving. A typical \MoE layer stores \(E\) experts but routes each token through only \(k \ll E\) of them, so the weights that a single decoding step touches are a small fraction of the layer's total parameters. 
However, in practice, existing edge serving engines~\citep{gerganov2023llamacpp,ktransformers2025,xue2024moeinfinity} deliver far less than what the hardware can theoretically support, and the shortfall is more severe on agentic workloads: time-to-first-token (\TTFT) grows with every tool call, and decode runs far below the machine's memory bandwidth.  This section examines the three challenges behind the shortfall: prefill costs (\sect{motivation-prefill}), decode costs (\sect{motivation-decode}), and the resource variability beneath both (\sect{motivation-resources}); the three subsections of \sect{design} answer them in order.

\subsection{Challenges in the Prefill Stage: Transfer and Recomputation Costs}
\lblsect{motivation-prefill}

Prefill determines \TTFT of every agent turn. On edge hardware, the prefill time mainly consists of two parts: expert transfer, which scales with the full model rather than the activated path, and context recomputation, which agentic sessions trigger far more often than current systems can afford.

\xpar{Expert transfer adds seconds on top of every prefill.}
Although decode routes each token to only \(k\) experts, the prefill process involves thousands of tokens per layer. As a result, these routed tokens activate nearly the entire expert set.  
A prefill pass therefore streams almost the complete expert pool through the \CPU--\GPU
interconnect. This introduces a severe I/O overhead that a \VRAM-resident deployment would not perform at all. Taking an FP4 deployment of DeepSeek-V4-Flash as an example, it requires transferring roughly 140\,GB of expert weights and adds roughly two seconds on an RTX 5090 system (PCIe~5.0~x16, \(\sim\)60\,GB/s), five seconds on RTX 4090- and 3090-class desktops (PCIe~4.0~x16, \(\sim\)25\,GB/s), and ten or more seconds on the x8 links common in laptops.  An engine that fetches experts on demand exposes this entire window as \GPU
idle time. This multi-second latency is usually not acceptable in agentic serving. 

\xpar{Agentic tool calls trigger frequent re-prefill.}
The second challenge is computation that repeats work the model has already done. Many frontier models adopt \HybridAttention architectures that interleave full attention with sliding-window attention (\eg, DeepSeek-V4-Flash~\citep{deepseekv4flash2026} and GPT-OSS~\citep{openai2025gptoss}) or recurrent layers (\eg, gated DeltaNet~\citep{yang2024gateddeltanet} in Qwen3.6-35B-A3B and Kimi Delta Attention~\citep{kimilinear2025} in Kimi-K3). Unlike standard attention, these layers compress the past context into a single state or a recent window of KV entries. Because each saved state consumes as much memory as the KV cache of hundreds of tokens, serving engines keep only a small number of checkpoints. 
Agentic workloads modify context on almost every turn. For example, tool calls often remove old outputs and delete thinking segments. Because of these modifications, any checkpoint taken after the modified position becomes invalid. The engine must fall back to the last valid checkpoint before the change. Because checkpoints are sparse, the engine often has to re-prefill thousands of tokens. Yet, consumer GPUs cannot hide this repeated cost: an RTX 5090 delivers roughly a fifth of an H100's and a tenth of a B200's dense BF16 throughput, so each redundant re-prefill of a long context occupies the \GPU for tens of seconds.

\subsection{Challenges in the Decode Stage: Cache Misses and Limited CPU Bandwidth}
\lblsect{motivation-decode}

Decode latency is determined by how the misses of each step's expert are served, and our analysis attributes the slowdown of existing engines to two hardware characteristics and one policy consequence: static expert placement misses most of the routed traffic, consumer \CPU bandwidth is too small to absorb the remainder alone, and the correct division of miss work between the two paths is hardware-specific.

\xpar{Static expert placement misses the routed traffic.}
Existing hybrid engines fix their expert placement at load or prefill time:
llama.cpp assigns \MoE tensors to devices when the model is
loaded~\citep{gerganov2023llamacpp}, and KTransformers pins a
``hot'' subset of experts in \GPU memory and executes the remainder
on the \CPU~\citep{ktransformers2025}. However, routing shifts with every token and every workload. A placement frozen at prefill time captures only a small fraction of the routed traffic. Consequently, the majority of expert evaluations fall to the \CPU, leaving both the \GPU and the \PCIe link idle (\sect{eval-breakdown}).

\xpar{Consumer CPUs cannot carry decode alone.}
At small decode batches, expert execution is memory-bound:
each token streams the routed expert weights once.  Consumer platforms attach the \CPU to two DRAM channels. This provides roughly 50\,GB/s of peak bandwidth for dual-channel DDR4 and 80--90\,GB/s for DDR5, compared to the 1--1.8\,TB/s a single RTX 4090 or 5090 draws from its on-package memory. Because of this bandwidth gap, a \CPU-only expert path caps decode speed at a small fraction of what the same weights would sustain from \VRAM, regardless of how many cores are available.

\xpar{The right division of work is hardware-specific.}
A missed expert can be transferred over \PCIe and executed on the
\GPU, or executed on the \CPU where its weights reside.  Neither approach is universally preferable. Relying on transfer alone leaves residual host bandwidth and \CPU cores idle whenever host memory can deliver more bytes than the link can move. Conversely, relying on \CPU execution alone leaves the \PCIe link idle and forfeits the future hits that a cache fill would buy.  The right mixture depends on the hardware. For example, an RTX 4060 laptop on LPDDR5 and an RTX 5090 desktop on DDR5 sit at opposite ends of the host-to-\PCIe balance. This optimal mixture cannot be read from specification sheets.  A fast design must therefore make this
division quantitatively, on the machine it actually runs on.

\subsection{Challenges in Resource Management: Nothing on the Edge Is Dedicated}
\lblsect{motivation-resources}

In datacenter environments, \GPU and \CPU resources are typically dedicated entirely to the serving workload. On edge devices like laptops and personal computers, however, LLM serving is just one of many concurrent applications. As a result, the resources available to the serving engine are highly dynamic.

\xpar{VRAM budget and its split change frequently during serving.}
On edge devices, the \GPU is shared with the desktop compositor, browsers, and games, each of which can claim gigabytes of \VRAM at any time. Therefore, the budget
available to a serving engine differs across launches and can shrink
or grow at any moment during serving.
The best split of that budget moves as well: agentic sessions
accumulate context across turns, so \KVCache demand grows while the
expert working set stays roughly fixed, and a split chosen on the
first turn is wrong many turns later.
\GPU memory must therefore stay adjustable at runtime, in total size
and in its split between the \KVCache and experts, without restarting
the engine.

\xpar{Engine startup is slow, and it happens often.}
Starting a serving engine is resource intensive: the complete expert pool must
be loaded from disk and the \GPU warmed before the first request.  For
the FP4 deployment of DeepSeek-V4-Flash, reading the roughly 140\,GB
pool from a 7\,GB/s NVMe drive alone takes about 20 seconds, before
any warmup begins.  On an edge machine this cost recurs: users open
the engine when they need it, close it to free the machine, and
switching to a different model restarts the engine as well.  An edge
engine must therefore start fast.

\section{FreeToken Design}
\lblsect{design}

\FreeToken organizes edge \MoE serving around a two-level expert-memory
hierarchy (\fig{overview}).  The \CpuExpertPool holds the complete routed-expert weights
and remains the source of truth, while non-expert weights stay resident
on the \GPU.  \FreeToken turns the remaining \GPU memory into a single
\ElasticGpuExpertCache shared by all \MoE layers: each slot holds every
tensor required to evaluate one layer–expert pair, so residency,
lookup, and execution all operate on logical
\((\text{layer},\text{expert})\) identifiers rather than tensor shards.

The design follows the two serving phases whose bottlenecks
\sect{motivation} identified.  During prefill
(\sect{design-prefill}), \FreeToken hides massive expert movement behind
computation and maintains prefixes (including recurrent states) that
survive agentic context editing.  During decode
(\sect{design-decode}), \BandwidthAdaptiveExecution divides cache
misses between \PCIe transfer and \CPU execution according to two
measured bandwidths of the machine.  Beneath both phases, an
elastic expert-memory lifecycle (\sect{design-lifecycle}) turns \GPU
cache capacity into a runtime-adjustable resource rather than a
loading-time constant.

\begin{figure}[t]
    \centering
    \includegraphics[width=\textwidth]{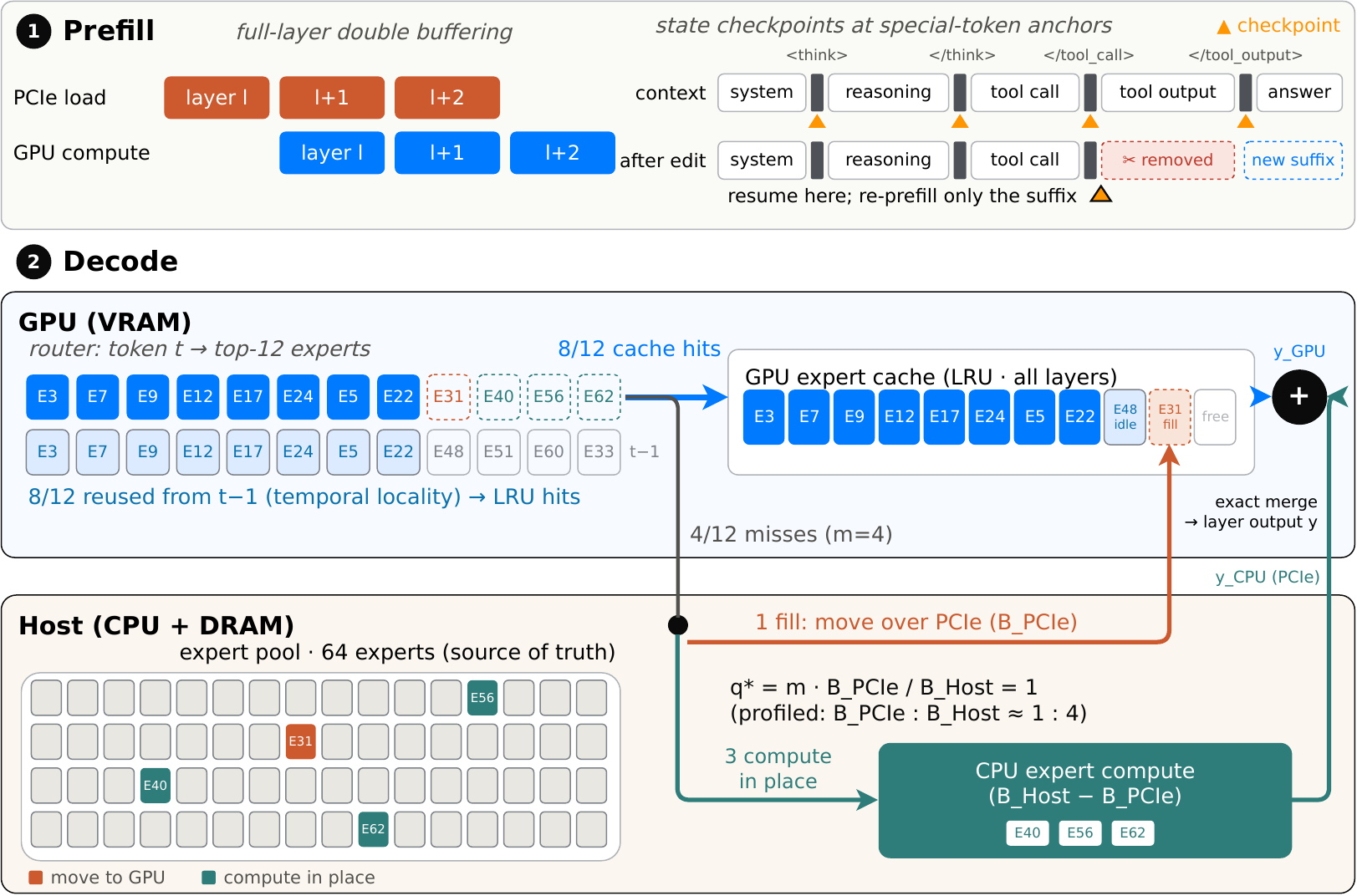}
    \caption{\FreeToken overview.  \textbf{(1)~Prefill:} expert
    loading is double-buffered at full-layer granularity, streaming
    layer \(l{+}1\) over \PCIe while the \GPU computes layer \(l\);
    recurrent-state checkpoints are anchored at special-token
    boundaries, so a context edit resumes from the nearest surviving
    anchor and re-prefills only the new suffix.
    \textbf{(2)~Decode:} most routed experts hit the shared LRU
    expert cache (here 8 of 12, following temporal locality).  The
    \(m{=}4\) misses are divided by \(q^\star = m\,\BPcie/\BHost\)
    between cache fills over \PCIe (one expert) and in-place \CPU
    execution (three), using bandwidths profiled on the deployed
    machine; the \GPU and \CPU partial outputs merge exactly.  The
    host-resident expert pool remains the source of truth
    throughout.}
    \lblfig{overview}
\end{figure}


\subsection{Prefill Codesign: Pipelined Loading and Semantic-Aware State Caching}
\lblsect{design-prefill}

\sect{motivation-prefill} attributed prefill cost to two components:
massive expert transfer and redundant recomputation.  \FreeToken answers
each with a dedicated mechanism: full-layer double buffering hides
the former behind computation, and the \SemanticAwareStateCache lets
prefixes, including recurrent state, survive the context edits that
would otherwise trigger the latter.

\xpar{Full-layer double buffering hides transfer behind computation.}
Because prefill activates nearly the entire expert set of every layer
(\sect{motivation-prefill}), \FreeToken does not fetch prefill experts
on demand.  It allocates two full-layer buffers from the global slot pool. While the \GPU computes the routed experts of layer \(l\) from one buffer, a dedicated transfer stream simultaneously loads the complete expert set of layer \(l+1\) into the other. By loading the full layer, the transfer can start before that layer's routing is even known. This ensures that weight movement proceeds continuously in the background, rather than serially between layers.  The buffers then swap roles.  Because they share the
decode cache's slot pool, there is no separate
prefill cache and no phase handoff, and entries that survive prefill
seed the latency-sensitive decode phase.  When the slot pool cannot
spare two full layers, \FreeToken falls back to on-demand prefill
loading rather than oversubscribing \GPU memory.

\xpar{Semantic anchors preserve recurrent state across context edits.}
\HybridAttention models add a second prefix resource beyond the
\KVCache.  \FreeToken manages full-attention KV with a radix prefix
tree, following existing serving systems~\citep{zheng2024sglang}; a
recurrent layer, however, compresses its entire prefix into one
evolving state that cannot be partially reused, so prefix reuse for
these layers depends on checkpoints of the state taken during prefill
and decode. To address this, \FreeToken maintains a \SemanticAwareStateCache. This is a small pool of recurrent-state checkpoints attached to nodes of the prefix tree. When a new request arrives, it restores from the deepest checkpoint whose position still survives after the prompt is edited.
Because each checkpoint captures the complete recurrent state of all
such layers, only a few can be held, and their placement determines
their value.  

\FreeToken spends this budget at \SemanticAnchors: the
special-token boundaries that mark thinking segments, tool calls
and outputs, and conversation turns. These are exactly the positions where
agent frameworks modify and truncate context: OpenClaw strips
thinking blocks from every assistant turn but the
latest~\citep{openclaw2026}; OpenCode replaces tool outputs beyond a
recent window with a fixed placeholder~\citep{opencode2026};
SWE-agent elides all but the last
\(n\) observations~\citep{yang2024sweagent}.
In every case, the edit replaces or removes \emph{whole blocks
marked by special tokens}, such as thinking segments, tool calls, tool outputs, and conversation turns.  Because the harness preserves the exact prefix up to the edited block, checkpoints taken at these semantic boundaries are far more likely to survive this truncation than those placed at arbitrary .  When an agent framework edits history after a tool call, the preserved prefix ends at such a boundary. A checkpoint anchored there lets full-attention layers reuse their \KVCache up to the edit point while recurrent layers resume from the anchor, so only the genuinely new suffix is re-prefilled.  Checkpoint
slots are recycled with LRU eviction, independently of the KV pool.

\subsection{Decode Codesign: Semantic-Aware Expert Caching and $q\star$ Policy}
\lblsect{design-decode}

At each \MoE layer during decode, the router and cache lookup run on
the \GPU and identify the active experts already resident in the
cache, \(\HitSet\), which execute on the \GPU directly.  The remaining challenge is how to serve the \(m\) unique missing experts \(\MissSet\).

\xpar{Semantic-aware expert caching follows the model's evolving computation.}
During decode, routing shows strong \TemporalExpertLocality: across consecutive steps, the same \MoE layer repeatedly routes to overlapping or recently used experts, a routing consistency measured across model families~\citep{liang2025locality}. \FreeToken turns this property into \GPU residency through semantic-aware expert caching. Rather than pinning experts according to a workload-agnostic placement chosen at load time, it maintains a shared LRU residency space whose contents continuously follow the experts selected by the router. A cache hit refreshes an expert's recency, a cache fill admits the newly selected expert, and eviction removes the expert least recently demanded by the model. In this way, scarce \GPU memory tracks the current working set of the generation. \sect{eval-breakdown} confirms this locality on our agentic traces and quantifies its advantage over static placement at equal cache capacity.

Caching cannot eliminate misses: cold starts, abrupt working-set shifts, and limited capacity still leave some selected experts off the \GPU. \BandwidthAdaptiveExecution serves these residual misses, deciding which experts to bring into the cache and which to execute in place on the \CPU.

\xpar{Measured bandwidths determine how misses are served.}
\FreeToken's \BandwidthAdaptiveExecution dynamically divides the \(m\) missing experts between \PCIe transfer and \CPU execution based on the balance of two measured bandwidths: the pinned expert-transfer bandwidth (\(\BPcie\)) and the host-side expert-processing bandwidth (\(\BHost\)). It splits the misses into a cache-fill set \(\FillSet\) and a \CPU-execution set \(\CpuSet\):
\begin{equation}
    \MissSet = \FillSet \,\dot{\cup}\, \CpuSet,
    \qquad
    q = |\FillSet|.
    \lbleqn{miss-partition}
\end{equation}
Experts in \(\FillSet\) are transferred into cache slots, execute on the
\GPU, and remain resident for future reuse; experts in \(\CpuSet\)
execute directly from the \CpuExpertPool and leave residency unchanged.
These two sets are served concurrently: cache fills
proceed at the full \PCIe rate, while the \CPU consumes only the host
bandwidth left over by the saturated link, converting residual host
bandwidth into current-token progress without suspending cache
updates.

The optimal split ratio is derived from a residual-bandwidth argument. Let \(S\) denote the size (in bytes) of one complete expert. Since both expert DMA transfers and \CPU execution read from the same host-memory subsystem, a saturated \PCIe transfer leaves a residual bandwidth of:
\begin{equation}
    \BResidual = \max(\BHost-\BPcie,\, 0)
    \lbleqn{residual-bandwidth}
\end{equation}
This residual bandwidth is precisely what is available for concurrent \CPU expert execution. Consequently, the execution times for the two branches are:
\begin{equation}
    T_{\mathrm{fill}}(q)
    \approx \frac{qS}{\BPcie},
    \qquad
    T_{\mathrm{cpu}}(m-q)
    \approx \frac{(m-q)S}{\BHost-\BPcie}.
    \lbleqn{branch-time}
\end{equation}
Balancing the concurrent branches gives
\begin{equation}
    \frac{q}{m-q}
    \approx
    \frac{\BPcie}{\BHost-\BPcie},
    \qquad
    q^\star
    \approx
    m\,\frac{\BPcie}{\BHost}.
    \lbleqn{dynamic-split}
\end{equation}
This single formulation covers all hardware balances. As \(\BHost\) approaches \(\BPcie\), \(q^\star\) approaches the total miss count \(m\), and the system degenerates into pure \OnDemandCacheFill without requiring separate execution branches or policies.

In practice, \FreeToken rounds \(q^\star\) to an integer and delegates the specific selection of \(\FillSet\) to the cache replacement policy. It always retains at least one fill, so the cache continues warming even when the \CPU handles most misses. The bandwidth parameters (\(\BHost\) and \(\BPcie\)) are empirically profiled on the target hardware at deployment. During execution, the \CPU and \GPU compute their respective partial sums and merge them, preserving the exact \MoE output without algorithmic approximation. In execution order, \FreeToken launches the \CPU branch first. It then runs the \GPU miss path, which consists of a cache update, a batch copy of \(\FillSet\), and a grouped evaluation of the combined \GPU execution set \(\GpuSet = \HitSet \cup \FillSet\). Concurrently, the \CPU workers process \(\CpuSet\). The exposed layer latency is therefore the slower of the two concurrent branches, which is precisely the quantity that \eqn{dynamic-split} balances. This per-layer control flow includes miss detection, set sizing, victim selection, and the \CPU branch itself. Keeping all of this inside a statically captured \CudaGraph is an
implementation problem of its own; \sect{implementation-kernels}
presents the graph-compatible cache and execution machinery.

\subsection{Elastic Memory Management for Edge-Native Runtimes}
\lblsect{design-lifecycle}

Unlike datacenter deployments, edge \GPU resources are rarely dedicated solely to LLM serving. The \VRAM available to a serving engine can fluctuate across launches and even change mid-session as the \GPU is shared with other applications. Furthermore, the memory demand itself shifts between the expert cache and the \KVCache as contexts grow, and engines are often started on demand. Consequently, building the \CpuExpertPool from disk is a user-visible cost. \FreeToken absorbs this variability with two mechanisms, both resting on one property:
because the \CpuExpertPool remains the source of truth, \GPU memory
affects only performance, never correctness.

\xpar{Runtime cache reconfiguration.}
After non-expert weights and runtime state are allocated, \FreeToken
divides the remaining \GPU-memory budget between \KVCache pages and
complete-expert slots. This division is not fixed at launch. At any scheduler safe point, \FreeToken can rebuild
the \GpuExpertCache for a revised \RuntimeVramBudget. This is done without restarting the engine or reloading the \CpuExpertPool, dynamically re-establishing the captured execution path for the new cache configuration.

\xpar{Fast engine bootstrap.}
Startup time has two components: loading the expert pool from disk
into host memory, and, traditionally, warming the \GPU before
serving.  \FreeToken shortens the first and eliminates the second.
Loading reads expert weights from disk directly into their final host
layout and pins the memory only afterward; pinning empty buffers
first would fault in and zero gigabytes of pages merely to overwrite
them.  Warming is unnecessary by construction: the first request is
served with a cold cache, its misses handled by the ordinary decode
path of \sect{design-decode}, and the cache heats up through normal
serving.

\section{Implementation}
\lblsect{implementation}

\FreeToken follows the GPU-centric serving architecture established by
systems such as SGLang and vLLM~\citep{zheng2024sglang,kwon2023vllm},
combining paged \KVCache management and radix-based prefix reuse with
community kernel libraries including FlashInfer~\citep{ye2025flashinfer}
and Flash Linear Attention~\citep{yang2024fla} where applicable.  On
this substrate, two implementation layers realize the design of
\sect{design}: the graph-compatible expert cache
(\sect{implementation-kernels}) and the storage and platform machinery
beneath it (\sect{implementation-runtime}).

\subsection{CUDA-Graph-Compatible LRU Cache}
\lblsect{implementation-kernels}

Expert caching is inherently dynamic: the specific experts that miss, the number of experts fetched, and the slots evicted all change at every step. Consequently, a host-controlled cache would reintroduce a costly device synchronization at every \MoE layer. To avoid this, \FreeToken keeps all routing-dependent control strictly on the \GPU. This dynamic control is represented as data inside a statically captured graph. Specifically, this data consists of fixed-shape work buffers and device-resident valid counts.

\xpar{Device-side cache control.}
For each \MoE layer, one \GPU kernel deduplicates the routed experts,
classifies them against the residency table, derives the
bandwidth-based fetch count \(q\), selects eviction victims, and
rewrites logical routed IDs into physical slot IDs or a special CPU-assignment flag. Victim selection avoids the classic LRU trap of requiring one full-cache scan per evicted slot. Instead, a single-pass kernel identifies the \(K\) least-recently-used candidate slots for eviction in one go. The miss path then simply consumes the first \(q\le K\) of these slots. As a result, victim discovery always costs exactly one pass, regardless of the realized miss count.  The resulting copy work list drives a single, fused transfer. Because every expert bank shares the same logical expert-to-slot mapping (\sect{implementation-runtime}), a single device-resident source/destination index list can be applied to all banks in one fixed-shape launch. A valid count is used to mask any unused work. This design yields few kernel launches, high \PCIe utilization, and removes routing-dependent decision overhead from the host.

\xpar{Graph-resident CPU execution.}
The \CPU branch of \BandwidthAdaptiveExecution is captured into the same graph. For each supported decode batch size, \FreeToken prepares stable pinned I/O buffers and persistent task descriptors. The device-to-host copies, a host-function submit node, the concurrent \GPU path, a synchronization node, and the host-to-device result copy are all captured together. As a result, replay re-executes the full heterogeneous step without requiring per-token Python scheduling.  The workers themselves form a persistent C++ pool pinned to physical cores. Their kernels consume expert weights using architecture-specific SIMD and in-kernel dequantization, which keeps the path bandwidth-bound. Finally, they return gate-weighted, per-token partial outputs.

\subsection{Expert Storage and Platform Adaptation}
\lblsect{implementation-runtime}

\xpar{Expert banks and the FTW format.}
\FreeToken normalizes model-specific checkpoint layouts into a small set of \ExpertBanks. Each bank uses the flattened layer--expert identifier \(lE+e\) as its leading dimension. Rows with the same identifier across all banks collectively form one complete expert. As a result, \GPU kernels and the \CPU executor share one logical expert identity, regardless of the underlying physical format.  To make loading fast, \FreeToken provides the FreeToken Weight (\FTW) format, which stores the expert weights merged into this runtime bank layout ahead of time. During engine launch, this format allows the system to skip tensor discovery and repacking entirely. Instead, it reads aligned chunks with parallel direct I/O straight into exact-size host banks, which are pinned only after they are filled (\sect{design-lifecycle}).

\xpar{Platform adaptation.}
At load time, \FreeToken selects \GPU kernels compatible with the expert representation, \GPU architecture, and CUDA environment. Meanwhile, the \CPU executor dispatches to the available SIMD implementation and physical-core layout.  When the complete expert pool cannot be pinned or registered for DMA (which is a restriction on some operating systems and driver configurations), \FreeToken falls back to a \PureCpuMoeBackend. Under this backend, expert weights stay in pageable host storage and all routed experts execute on the \CPU. Non-expert layers remain on the \GPU, and only activation-sized inputs, routing metadata, and aggregated outputs cross the CPU--GPU boundary.  This path trades peak transfer bandwidth
for deployability on platforms where the fast path cannot be
established.

\section{Evaluation}
\lblsect{evaluation}

We evaluate \FreeToken on real agentic workloads, serving \MoE models
whose complete expert pools exceed \VRAM, against actively maintained
edge serving engines on six machines.  \sect{eval-setup} details the
setup, \sect{eval-e2e} reports the main end-to-end results, and
\sect{eval-breakdown} attributes the gains to \FreeToken's mechanisms
and tests their generality across hardware.

\subsection{Experimental Setup}
\lblsect{eval-setup}

\xpar{Hardware.}
Six discrete-\GPU systems (\tbl{hardware}): five consumer machines
spanning the effective host and \PCIe bandwidth range of current edge
hardware, and one workstation-class box (a single RTX PRO 6000
Blackwell, 96\,GB) hosting the frontier-scale demonstration.  The
3090, 4090, and 5090 systems are rented dual-socket servers whose
\CPU{}s far exceed any edge host, so every serving run and bandwidth
measurement on them is capped at 6 \CPU threads and pinned to the
\GPU's NUMA node.  Capped this way, the servers deliver
56.7--77.3\,GB/s of host bandwidth, the same scale the two real edge
machines reach at their natural full threads (53.8\,GB/s on the
desktop's 16 cores, 47.5\,GB/s on the laptop's 14); the desktop and
laptop run untiered and validate the emulation on real edge hardware.
All bandwidths in \tbl{hardware} are measured on the deployed tensor
shapes rather than taken from platform specifications.

\begin{table}[t]
    \centering
    \caption{Test systems.  \(\BPcie\) is the measured
    host-to-device expert-transfer bandwidth over \PCIe;
    \(\BHost\) is the measured effective bandwidth of the
    \CPU-side \MoE expert kernel.  On the
    three rented servers the CPU-thread and DRAM columns give
    container quotas.}
    \lbltbl{hardware}
    \small
    \setlength{\tabcolsep}{3.5pt}
    \begin{tabular}{llcccrr}
        \toprule
        System & GPU (VRAM) & \PCIe & \(\BPcie\) & CPU (threads) & DRAM & \(\BHost\) \\
        & & & (GB/s) & & (GiB) & (GB/s) \\
        \midrule
        5090 & RTX 5090 (32\,GB) & 5.0 \(\times\)16 & 52.7 & 2\(\times\) Xeon Gold 6459C (32) & DDR5\, 180 & 77.3 \\
        4090 & RTX 4090 (24\,GB) & 4.0 \(\times\)16 & 25.1 & 2\(\times\) Xeon Platinum 8358P (32) & DDR4\, 240 & 63.2 \\
        3090 & RTX 3090 (24\,GB) & 4.0 \(\times\)16 & 25.3 & 2\(\times\) Xeon Gold 6330 (28) & DDR4\, 180 & 56.7 \\
        5090 desktop & RTX 5090 (32\,GB) & 5.0 \(\times\)16 & 49.0 & Ryzen 9 9950X3D (32) & DDR5\, 192 & 53.8 \\
        4060 laptop & RTX 4060 Laptop (8\,GB) & 4.0 \(\times\)8 & 11.8 & Core i9-13900H (20) & LPDDR5\, 32 & 47.5 \\
        PRO 6000 & RTX PRO 6000 (96\,GB) & 5.0 \(\times\)16 & 51.5 & Xeon Platinum 8559C (48) & DDR5\, 512 & 178 \\
        \bottomrule
    \end{tabular}
\end{table}

\xpar{Models.}
Two \MoE models: DeepSeek-V4-Flash (284B parameters, 13B
active)~\citep{deepseekv4flash2026}, served from the official
checkpoint whose routed experts are natively MXFP4-quantized, and
Qwen3.6-35B-A3B~\citep{qwen36a3b2026} in BF16, which gives exact
precision parity across engines (the 8\,GB laptop serves its official
NVFP4 release~\citep{qwen36nvfp42026} instead).  The cross-hardware
study adds GLM-5.2~\citep{glm522026} (753B parameters, 40B active;
NVFP4 routed experts~\citep{glm52nvfp42026}, a 433\,GB checkpoint),
served by the RTX PRO 6000 as a frontier-scale tier.

\xpar{Workloads.}
Four agentic scenarios.
\emph{W1, math reasoning}: AIME competition problems answered with
long chain-of-thought decoding and no tool use---single-turn and
decode-dominated.  \emph{W2, coding agent}: a SWE-bench repository
issue solved through the OpenCode harness with real tool execution
over three scripted user turns.  \emph{W3, coding agent, native
protocol}: the same issue driven by Claude Code through each engine's
Anthropic-compatible endpoint, which spawns concurrently-requesting
subagents and grows sessions to 56--65k tokens.  \emph{W4,
email/calendar agent}: thirteen fixed user turns over a mailbox kit
through OpenClaw at stock configuration (its 120\,s idle watchdog is
disabled so that slow engines remain measurable), carrying a
\(\sim\)24.5k-token system-context floor.  All engines serve the same
harness with identical requests; the coding runs must produce
the reference gold patch, and the W4 runs must complete all thirteen
turns.

\xpar{Baselines.}
llama.cpp, Ollama, KTransformers, and
MoE-Infinity~\citep{xue2024moeinfinity} on the configurations it
supports.  Weight formats are exactly aligned: every engine serves
Qwen3.6 in BF16, and every engine consumes DSV4-Flash's native MXFP4
expert blocks bit-exactly.

\xpar{Metrics.}
Decode throughput (per-request mean tok/s) and \TTFT (per-request
mean).  Agent trajectories diverge across
engines so cross-engine wall-clock totals are not compared.

\subsection{Main Results}
\lblsect{eval-e2e}

\begin{figure*}[t]
    \centering
    \includegraphics[width=\textwidth]{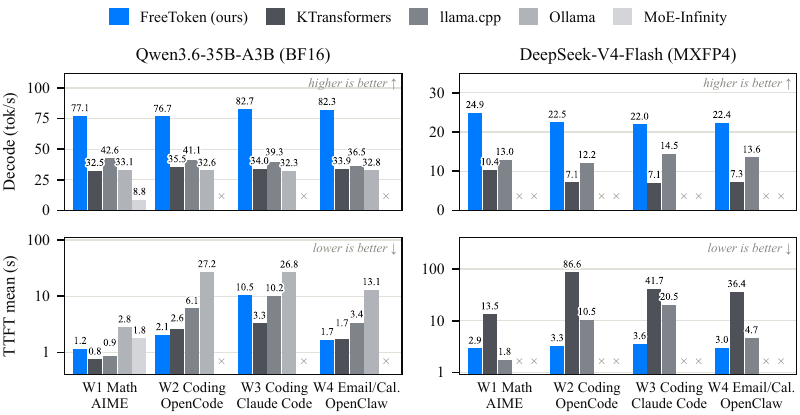}
    \caption{End-to-end serving on the RTX~5090 across four
    workloads (1. AIME, 2. OpenCode+SWE, 3. Claude Code+SWE, 4.OpenClaw+Email/Cal) and two models (Qwen3.6-35B-A3B BF16 and DeepSeek-V4-Flash MXFP4). 
    Top: decode TPS; bottom: mean TTFT (log scale).
    $\times$ marks configurations an
    engine cannot serve (Ollama and MoE-Infinity lack DSV4 support;
    MoE-Infinity provides no usable server for multi-turn agents).}
    \lblfig{exp1-main}
\end{figure*}

\fig{exp1-main} reports decode throughput and mean \TTFT for every
engine on both models across the four workloads (RTX 5090; Qwen3.6 at
6 \CPU threads, DSV4-Flash at 8).

\xpar{Decode throughput.}
\FreeToken sustains 77--83\,tok/s on Qwen3.6 and 22--25\,tok/s on
DSV4-Flash, 1.8--2.3\(\times\) and 1.5--1.9\(\times\) the strongest
baseline in each workload.  The rate is also stable under agentic
serving: it stays within 12\% of the single-turn W1 value across the
three agent workloads, while the most context-sensitive baseline,
KTransformers on DSV4-Flash, has already lost 31\% of its W1 rate at
W2.  Single-stream benchmarks therefore overstate baseline agentic
performance.  MoE-Infinity serves only W1 (8.8\,tok/s): its per-expert
prefill staging cap aborts the longer-prompt workloads, and its
bundled server retains no \KVCache across requests.

\xpar{Time to first token.}
\FreeToken posts the lowest mean \TTFT in five of the six multi-turn
cells (Qwen3.6\(\times\)W3 favors KTransformers' \GPU-prefill arm),
while W1's short isolated prompts favor llama.cpp.  The tails
separate the engines more sharply than the means: \FreeToken's worst
turn stays below 44\,s in every cell, while each baseline crosses
150\,s somewhere---llama.cpp at 232\,s, Ollama at 179\,s,
KTransformers at 946\,s.  These stalls cross the thresholds at which
real clients abandon the request: OpenClaw ships a 120\,s idle
watchdog, and Claude Code's default request timeout is roughly ten
minutes.  Tail \TTFT is therefore an availability boundary, not a
latency statistic.

\subsection{Breakdown and Cross-Hardware Analysis}
\lblsect{eval-breakdown}

\begin{figure}[t]
    \centering
    \includegraphics[width=\textwidth]{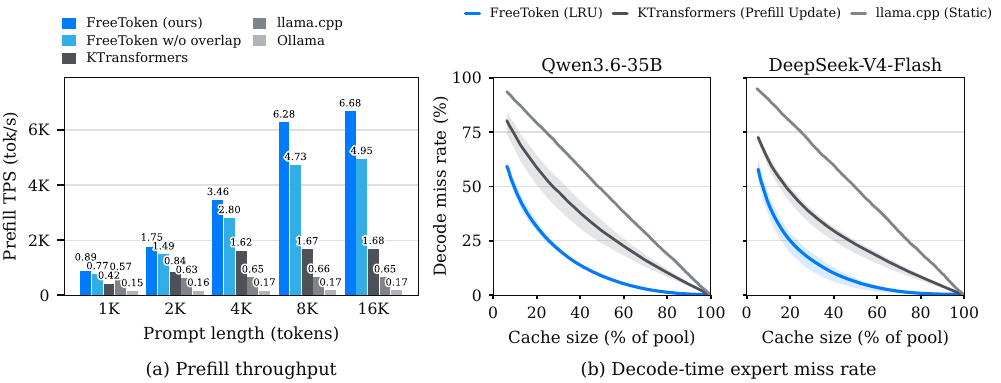}
    \caption{(a) Prefill TPS versus prompt length (RTX~5090,
    Qwen3.6-35B BF16), with and without \FreeToken's pipelined
    full-layer loading.      (b) Decode-time expert miss rate versus
    cache size (as a percentage of the expert pool) under the three
    engines' placement policies, replayed on identical routing
    traces; lines are means over W1--W4, bands the min--max range.}
    \lblfig{prefill-locality}
\end{figure}

\begin{figure}[t]
    \centering
    \includegraphics[width=0.70\textwidth]{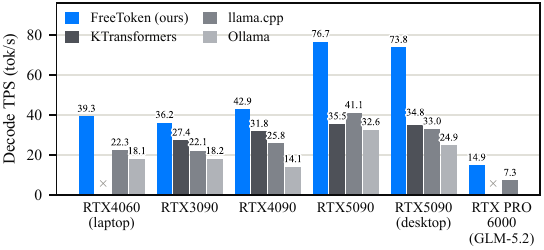}
    \caption{Coding-agent decode TPS across consumer GPUs (SWE issues via the OpenCode harness), Qwen3.6-35B-A3B. 4060 laptop using NVFP4, the other Qwen3.6 columns BF16. The RTX PRO 6000 column is a separate demonstration: GLM-5.2 (753B-A40B, NVFP4) on the math workload; Ollama is not run there. $\times$ marks configurations an engine cannot serve. }
    \lblfig{exp3-crosshw}
\end{figure}

Three analyses attribute the end-to-end gains to \FreeToken's
mechanisms and test their generality: pipelined prefill
(\fig{prefill-locality}a), expert-cache locality
(\fig{prefill-locality}b), and serving across the full hardware range
(\fig{exp3-crosshw}).

\xpar{Pipelined prefill.}
Full-layer double buffering makes prefill transfer-bound: with
overlap on, each 8,192-token prefill chunk completes in
1.19--1.22\,s, the time to stream the 64.4\,GB expert pool once at
52.7\,GB/s---the practical ceiling of the \PCIe~5.0 \(\times\)16
link---so expert computation is fully hidden behind transfer and
throughput climbs to 6.7k\,tok/s at 16k tokens
(\fig{prefill-locality}a).  Disabling the second buffer serializes
transfer against computation and costs 19\% of throughput at 4k
tokens, 25\% at 8k, and 26\% at 16k, the penalty growing with prompt
length as the share of hidden computation rises.

\xpar{Expert locality.}
Decode routing carries enough short-range locality that a per-miss
LRU beats placement chosen at prefill time, the observation
\sect{design-decode} builds on.  \fig{prefill-locality}b replays
identical routing traces from all four workloads against the three
engines' placement policies at equal cache capacity.  At the RTX 5090
serving capacity (37\% of Qwen3.6's expert pool, 11\% of
DSV4-Flash's), \FreeToken's global LRU misses 16\% and 39\% of
decode-time expert reads, against 41\% and 59\% for KTransformers'
prefill-updated placement and 62\% and 89\% for
llama.cpp's routing-blind static split.  The ordering holds across
workloads at every capacity short of the full pool.

\xpar{Cross-hardware serving.}
The advantage holds on every machine.  \fig{exp3-crosshw} repeats W2
across the five consumer systems: \FreeToken leads the strongest
baseline by 1.3\(\times\) on the RTX 3090 and 4090, 1.9\(\times\) on
the 5090 server, 2.1\(\times\) on the 5090 desktop, and 1.8\(\times\)
on the RTX 4060 laptop, where the NVFP4 build sustains 39.3\,tok/s on
an 8\,GB, \PCIe \(\times\)8 machine---92\% of the RTX 4090 rate.  The
two 5090 columns share the same \GPU silicon and differ only in the
host: moving from the many-channel server to a dual-channel consumer
desktop costs \FreeToken 4\% of its decode rate, while llama.cpp
keeps only 80\% of its rate as its \CPU-resident experts starve on
two DDR5 channels.  At the frontier tier, \FreeToken serves GLM-5.2
on the single RTX PRO 6000 at 14.9\,tok/s against llama.cpp's 7.3
(2.0\(\times\)) with bit-identical expert weights and comparable mean
\TTFT (7.5 versus 7.8\,s).  KTransformers has no servable path for
this model on the box: its GLM-5.2 methods require 753\,GB--1.5\,TB
of host-resident experts against 512\,GiB of host memory, and its
\CPU kernels do not read GLM-5.2's NVFP4 layout.


\section{Related Work}
\lblsect{related-work}

\xpar{Expert offloading and caching.}
Serving \MoE models whose expert pool exceeds \GPU memory has converged
on the architecture \FreeToken also adopts: the complete expert pool
resides in host memory or on disk, and a subset of experts is cached on
the \GPU.  EdgeMoE~\citep{yi2023edgemoe} established this design for
on-device inference; Mixtral-offloading~\citep{eliseev2023fastmoe}
combined an LRU expert cache with speculative prefetching;
MoE-Infinity~\citep{xue2024moeinfinity} traces request-level activation
patterns to guide prefetching and caching; and
ProMoE~\citep{song2024promoe}, ExpertFlow~\citep{he2024expertflow}, and
FineMoE~\citep{fan2025finemoe} sharpen the predictors that anticipate
future routing.  Routing-consistency measurements across \MoE model
families~\citep{liang2025locality,lin2025expertcaching} support the
premise that such caches can be effective.  These systems differ in how
well they predict misses, but not in how they serve them: every miss is
ultimately a \PCIe transfer, so decode latency remains bounded by the
link no matter how accurate prediction becomes, while host compute
capacity sits idle.  A complementary line lowers the transfer volume by
relaxing fidelity: HOBBIT~\citep{tang2024hobbit} fetches
reduced-precision replicas of missed experts, and
SiDA~\citep{du2024sida} and SMoE~\citep{zhu2025smoe} substitute or skip
low-scoring experts (trading accuracy for bandwidth), whereas Pre-gated
MoE~\citep{hwang2024pregated} restructures and fine-tunes the router
itself.  \FreeToken keeps the routed computation exact and the model
unmodified; it changes how residual misses are served rather than how
well they are predicted.

\xpar{Hybrid CPU--GPU execution.}
A second line enlists the \CPU as a compute resource rather than a
weight store.  For dense models, FlexGen~\citep{sheng2023flexgen} and
DeepSpeed-Inference~\citep{aminabadi2022deepspeed} stream weights at
layer granularity for throughput-oriented batch inference, and
PowerInfer~\citep{song2023powerinfer} splits neurons by activation
statistics, which requires ReLU-family sparsity and learned predictors;
llama.cpp and Ollama~\citep{gerganov2023llamacpp,ollama2023} assign
whole layers to devices statically at load time.  For \MoE,
Fiddler~\citep{kamahori2025fiddler} first treated a missed expert as
work that can execute on the \CPU rather than only data to move;
KTransformers~\citep{ktransformers2025} made in-place \CPU expert
execution fast with AMX-optimized kernels;
HybriMoE~\citep{zhong2025hybrimoe} rebalances \CPU/\GPU queues with
per-step schedule simulation; SMoE~\citep{zhu2025smoe} balances loading
against \CPU time with a greedy two-pointer heuristic; and other
systems execute every miss on the \CPU~\citep{huang2025collaborative}
or pipeline \CPU, \GPU, and I/O for offline
throughput~\citep{cao2024moelightning,klotski2025}.  Across this line,
the division of work is either fixed at startup (KTransformers keeps
routed experts on the \CPU even when \PCIe is idle and cache capacity
is available) or recomputed by host-side heuristics whose scheduling
cost and per-layer synchronization cannot be captured into a
\CudaGraph; llama.cpp likewise cannot maintain graph execution in its
hybrid mode.  Moreover, these systems are designed and evaluated
around single-shot, short-prompt inference, often on unquantized
weights; none provides the cross-request prefix reuse that multi-turn
agentic sessions re-enter prefill with at every tool-calling turn.
\FreeToken instead derives the division from two measured bandwidths as
a closed-form ratio (cheap enough to remain device-resident inside a captured graph) and embeds it in a serving runtime rather than a
single-request harness.

\xpar{Serving infrastructure and hierarchical memory.}
\FreeToken builds on the \GPU-centric serving substrate of
vLLM~\citep{kwon2023vllm} and SGLang~\citep{zheng2024sglang}, and on
FlashInfer~\citep{ye2025flashinfer}, whose attention scheduling set a
precedent for dynamic behavior inside \CudaGraph capture.  On the
memory-management side, SGLang HiCache~\citep{hicache2025} tiers the
\KVCache across \GPU, host, and remote storage to preserve prefix reuse
in long, multi-turn sessions; WiSP~\citep{wisp2026} splits \VRAM
between expert weights and \KVCache by marginal latency value;
eLLM~\citep{ellm2025} rebalances elastic memory pools at runtime; and
FluxMoE~\citep{fluxmoe2026} pages experts to prioritize \KVCache
capacity.  These managers move passive bytes: when a page or expert is
absent, it can only be fetched, and WiSP's own analysis finds
single-stream decode \PCIe-bound regardless of prediction
accuracy, which is the ceiling that allocation alone cannot lift.  Expert
weights offer one further degree of freedom, since a missing expert can
also be computed where it resides.  \FreeToken combines both levers: it
applies the hierarchical-memory philosophy to the expert pool. It uses an
elastic full-expert cache with unified prefill/decode
residency, and adds bandwidth-adaptive \CPU co-execution on the same
serving substrate.  Our contribution is the
integrated runtime and the measured-bandwidth model that coordinates
caching, transfer, and \CPU execution within it.

\section{Conclusion}
\lblsect{conclusion}

We presented \FreeToken, an edge-native serving system for frontier-scale \MoE models on personal hardware. FreeToken is built around a simple observation: once sparse activation makes the computation of a model feasible, local inference becomes less a question of whether the model fits on a GPU and more a question of how well the system can orchestrate the machine. FreeToken therefore treats the GPU, CPU, host memory, and interconnect as a unified inference platform, adapting model state and execution to the changing structure of agentic workloads and the hardware on which they run. FreeToken spans more than 20 \MoE models and hardware from an 8 GB laptop GPU to a workstation GPU, enabling models from 35B to 753B parameters while consistently outperforming existing edge serving systems. More broadly, our results suggest that the boundary of local AI is increasingly determined not only by hardware capacity but also by the serving software that composes the resources already available. \FreeToken takes a step toward turning open weights into open access, making personal machines a practical platform for frontier-scale intelligence.

\section*{Acknowledgements}
We thank Jiaming Tang and Ganxiang Yang for testing the system, and
Tian Xia for assembling the RTX 5090 desktop used in our experiments.

\bibliographystyle{plainnat}
\bibliography{paper}

\begin{thebibliography}{49}
\providecommand{\natexlab}[1]{#1}
\providecommand{\url}[1]{\texttt{#1}}
\expandafter\ifx\csname urlstyle\endcsname\relax
  \providecommand{\doi}[1]{doi: #1}\else
  \providecommand{\doi}{doi: \begingroup \urlstyle{rm}\Url}\fi

\bibitem[Aminabadi et~al.(2022)]{aminabadi2022deepspeed}
Reza~Yazdani Aminabadi et~al.
\newblock Deepspeed inference: Enabling efficient inference of transformer models at unprecedented scale, 2022.
\newblock \url{https://arxiv.org/abs/2207.00032}.

\bibitem[{Anthropic}(2026)]{claudecodecosts2026}
{Anthropic}.
\newblock Manage costs effectively --- claude code documentation.
\newblock \url{https://code.claude.com/docs/en/costs}, 2026.
\newblock Average \$13 per developer per active day and \$150--250 per developer per month across enterprise deployments.

\bibitem[Cao et~al.(2024)Cao, Liu, Griggs, Schafhalter, Liu, Sheng, Gonzalez, Zaharia, and Stoica]{cao2024moelightning}
Shiyi Cao, Shu Liu, Tyler Griggs, Peter Schafhalter, Xiaoxuan Liu, Ying Sheng, Joseph~E. Gonzalez, Matei Zaharia, and Ion Stoica.
\newblock Moe-lightning: High-throughput moe inference on memory-constrained gpus, 2024.
\newblock \url{https://arxiv.org/abs/2411.11217}.

\bibitem[{DeepSeek-AI}(2026)]{deepseekv4flash2026}
{DeepSeek-AI}.
\newblock {DeepSeek-V4-Flash-0731}.
\newblock Model checkpoint, \url{https://huggingface.co/deepseek-ai/DeepSeek-V4-Flash-0731}, 2026.

\bibitem[Du et~al.(2024)Du, Li, Wu, Jiang, Sun, Zheng, Wu, Li, Li, and Chen]{du2024sida}
Zhixu Du, Shiyu Li, Yuhao Wu, Xiangyu Jiang, Jingwei Sun, Qilin Zheng, Yongkai Wu, Ang Li, Hai Li, and Yiran Chen.
\newblock Sida: Sparsity-inspired data-aware serving for efficient and scalable large mixture-of-experts models, 2024.
\newblock \url{https://arxiv.org/abs/2310.18859}.

\bibitem[Eliseev and Mazur(2023)]{eliseev2023fastmoe}
Artyom Eliseev and Denis Mazur.
\newblock Fast inference of mixture-of-experts language models with offloading, 2023.
\newblock \url{https://arxiv.org/abs/2312.17238}.

\bibitem[{eLLM authors}(2025)]{ellm2025}
{eLLM authors}.
\newblock ellm: Elastic memory management for efficient llm serving, 2025.
\newblock \url{https://arxiv.org/abs/2506.15155}.

\bibitem[{Fan et~al.}(2025)]{fan2025finemoe}
{Fan et~al.}
\newblock Taming latency-memory trade-off in moe-based llm serving via fine-grained expert offloading, 2025.
\newblock \url{https://arxiv.org/abs/2502.05370}.

\bibitem[{Fang et~al.}(2025)]{klotski2025}
{Fang et~al.}
\newblock Klotski: Efficient mixture-of-expert inference via expert-aware multi-batch pipeline, 2025.
\newblock \url{https://arxiv.org/abs/2502.06888}.

\bibitem[{FluxMoE authors}(2026)]{fluxmoe2026}
{FluxMoE authors}.
\newblock Fluxmoe: Expert paging for memory-efficient moe serving, 2026.
\newblock \url{https://arxiv.org/abs/2604.02715}.

\bibitem[Gerganov et~al.(2023)]{gerganov2023llamacpp}
Georgi Gerganov et~al.
\newblock llama.cpp: Llm inference in c/c++.
\newblock Software, 2023.
\newblock \url{https://github.com/ggml-org/llama.cpp}.

\bibitem[{He et~al.}(2024)]{he2024expertflow}
{He et~al.}
\newblock Expertflow: Optimized expert activation and token allocation for efficient mixture-of-experts inference, 2024.
\newblock \url{https://arxiv.org/abs/2410.17954}.

\bibitem[{Huang et~al.}(2025)]{huang2025collaborative}
{Huang et~al.}
\newblock Efficient cpu-gpu collaborative inference for moe large language models, 2025.
\newblock \url{https://arxiv.org/abs/2512.16473}.

\bibitem[Hwang et~al.(2024)Hwang, Wei, Cao, Hwang, Tang, Cao, and Yang]{hwang2024pregated}
Ranggi Hwang, Jianyu Wei, Shijie Cao, Changho Hwang, Xiaohu Tang, Ting Cao, and Mao Yang.
\newblock Pre-gated moe: An algorithm-system co-design for fast and scalable mixture-of-expert inference, 2024.
\newblock \url{https://arxiv.org/abs/2308.12066}.

\bibitem[Kamahori et~al.(2024)Kamahori, Tang, Gu, Zhu, and Kasikci]{kamahori2025fiddler}
Keisuke Kamahori, Tian Tang, Yile Gu, Kan Zhu, and Baris Kasikci.
\newblock Fiddler: Cpu-gpu orchestration for fast inference of mixture-of-experts models, 2024.
\newblock \url{https://arxiv.org/abs/2402.07033}.

\bibitem[{Kimi Team}(2025)]{kimilinear2025}
{Kimi Team}.
\newblock {Kimi Linear}: An expressive, efficient attention architecture, 2025.
\newblock \url{https://arxiv.org/abs/2510.26692}.

\bibitem[{Kimi Team}(2026)]{kimik32026}
{Kimi Team}.
\newblock {Kimi K3: Open Frontier Intelligence}, 2026.
\newblock \url{https://arxiv.org/abs/2607.24653}.

\bibitem[{KVCache-AI Team}(2025)]{ktransformers2025}
{KVCache-AI Team}.
\newblock Ktransformers: Unleashing the full potential of cpu/gpu hybrid inference for moe models.
\newblock SOSP 2025; software at \url{https://github.com/kvcache-ai/ktransformers}, 2025.

\bibitem[Kwon et~al.(2023)Kwon, Li, Zhuang, Sheng, Zheng, Yu, Gonzalez, Zhang, and Stoica]{kwon2023vllm}
Woosuk Kwon, Zhuohan Li, Siyuan Zhuang, Ying Sheng, Lianmin Zheng, Cody~Hao Yu, Joseph~E. Gonzalez, Hao Zhang, and Ion Stoica.
\newblock Efficient memory management for large language model serving with pagedattention, 2023.
\newblock \url{https://arxiv.org/abs/2309.06180}.

\bibitem[{Liang et~al.}(2025)]{liang2025locality}
{Liang et~al.}
\newblock Local routing consistency of mixture-of-experts language models, 2025.
\newblock \url{https://arxiv.org/abs/2505.16056}.

\bibitem[{Lin et~al.}(2025)]{lin2025expertcaching}
{Lin et~al.}
\newblock An in-depth study of llm serving with expert caching and prefetching for mixture-of-experts models, 2025.
\newblock \url{https://arxiv.org/abs/2511.05814}.

\bibitem[{LMArena}(2026)]{codearena2026}
{LMArena}.
\newblock Code arena: Webdev leaderboard.
\newblock \url{https://arena.ai/leaderboard/code/webdev}, 2026.
\newblock Snapshot of 2026-08-10.

\bibitem[{NVIDIA}(2026{\natexlab{a}})]{glm52nvfp42026}
{NVIDIA}.
\newblock {GLM-5.2-NVFP4}.
\newblock NVFP4 model checkpoint, \url{https://huggingface.co/nvidia/GLM-5.2-NVFP4}, 2026{\natexlab{a}}.

\bibitem[{NVIDIA}(2026{\natexlab{b}})]{qwen36nvfp42026}
{NVIDIA}.
\newblock {Qwen3.6-35B-A3B-NVFP4}.
\newblock NVFP4 model checkpoint, \url{https://huggingface.co/nvidia/Qwen3.6-35B-A3B-NVFP4}, 2026{\natexlab{b}}.

\bibitem[{Ollama Team}(2023)]{ollama2023}
{Ollama Team}.
\newblock Ollama.
\newblock Software, 2023.
\newblock \url{https://github.com/ollama/ollama}.

\bibitem[{OpenAI}(2025)]{openai2025gptoss}
{OpenAI}.
\newblock {gpt-oss-120b} \& {gpt-oss-20b} model card, 2025.
\newblock \url{https://arxiv.org/abs/2508.10925}.

\bibitem[{OpenClaw Contributors}(2026)]{openclaw2026}
{OpenClaw Contributors}.
\newblock {OpenClaw}: A personal ai agent platform.
\newblock \url{https://github.com/openclaw/openclaw}, 2026.
\newblock \texttt{dropThinkingBlocks} in the embedded agent runner strips thinking blocks from all assistant turns except the latest.

\bibitem[{OpenCode}(2026)]{opencode2026}
{OpenCode}.
\newblock Compaction --- opencode documentation.
\newblock \url{https://opencode.ai/v2/docs/compaction}, 2026.
\newblock Pruning replaces tool outputs older than a protected recent window with a fixed truncation placeholder.

\bibitem[{Presenc AI Research}(2026)]{gpusupply2026}
{Presenc AI Research}.
\newblock Gpu shipment tracker: Blackwell to rubin 2026.
\newblock \url{https://presenc.ai/research/gpu-shipment-tracker-blackwell-rubin-2026}, 2026.
\newblock Rubin 2026 supply projected at 200--300k units against a 5.7M target; hyperscalers capture 60--70\% of first-year supply.

\bibitem[{Qwen Team}(2026)]{qwen36a3b2026}
{Qwen Team}.
\newblock {Qwen3.6-35B-A3B}.
\newblock Model checkpoint, \url{https://huggingface.co/Qwen/Qwen3.6-35B-A3B}, 2026.

\bibitem[{SGLang Team}(2025)]{hicache2025}
{SGLang Team}.
\newblock Sglang hicache: Fast hierarchical kv caching with your favorite storage backends.
\newblock LMSYS Org blog, 2025.
\newblock \url{https://lmsys.org/blog/2025-09-10-sglang-hicache/}.

\bibitem[Sheng et~al.(2023)Sheng, Zheng, Yuan, Li, Ryabinin, Chen, Liang, R{\'e}, Stoica, and Zhang]{sheng2023flexgen}
Ying Sheng, Lianmin Zheng, Binhang Yuan, Zhuohan Li, Max Ryabinin, Beidi Chen, Percy Liang, Christopher R{\'e}, Ion Stoica, and Ce~Zhang.
\newblock Flexgen: High-throughput generative inference of large language models with a single gpu, 2023.
\newblock \url{https://arxiv.org/abs/2303.06865}.

\bibitem[{Simon Carless (GameDiscoverCo), reported by gHacks}(2026)]{steammau2026}
{Simon Carless (GameDiscoverCo), reported by gHacks}.
\newblock Steam passes 200 million monthly active users.
\newblock \url{https://www.ghacks.net/2026/07/12/steam-passes-200-million-monthly-active-users-50-more-than-playstation/}, 2026.

\bibitem[Song et~al.(2024)Song, Zhong, Chen, and Chen]{song2024promoe}
Xiaoniu Song, Zihang Zhong, Rong Chen, and Haibo Chen.
\newblock Promoe: Fast moe-based llm serving using proactive caching, 2024.
\newblock \url{https://arxiv.org/abs/2410.22134}.

\bibitem[Song et~al.(2023)Song, Mi, Xie, and Chen]{song2023powerinfer}
Yixin Song, Zeyu Mi, Haotong Xie, and Haibo Chen.
\newblock Powerinfer: Fast large language model serving with a consumer-grade gpu, 2023.
\newblock \url{https://arxiv.org/abs/2312.12456}.

\bibitem[Tang et~al.(2024)]{tang2024hobbit}
Peng Tang et~al.
\newblock Hobbit: A mixed precision expert offloading system for fast moe inference, 2024.
\newblock \url{https://arxiv.org/abs/2411.01433}.

\bibitem[{Valve Corporation}(2026)]{steamsurvey2026}
{Valve Corporation}.
\newblock Steam hardware \& software survey.
\newblock \url{https://store.steampowered.com/hwsurvey}, 2026.
\newblock June 2026: NVIDIA GPUs in $\sim$72\% of surveyed systems; RTX 4060 Laptop GPU most common at 3.81\%.

\bibitem[{WiSP authors}(2026)]{wisp2026}
{WiSP authors}.
\newblock Wisp: A working-set view of moe serving with bandwidth-aware expert paging, 2026.
\newblock \url{https://arxiv.org/abs/2606.21868}.

\bibitem[Xue et~al.(2024)Xue, Fu, Lu, Mai, and Marina]{xue2024moeinfinity}
Leyang Xue, Yao Fu, Zhan Lu, Luo Mai, and Mahesh Marina.
\newblock Moe-infinity: Efficient moe inference on personal machines with sparsity-aware expert cache, 2024.
\newblock \url{https://arxiv.org/abs/2401.14361}.

\bibitem[Yang et~al.(2024{\natexlab{a}})Yang, Jimenez, Wettig, Lieret, Yao, Narasimhan, and Press]{yang2024sweagent}
John Yang, Carlos~E. Jimenez, Alexander Wettig, Kilian Lieret, Shunyu Yao, Karthik Narasimhan, and Ofir Press.
\newblock {SWE-agent}: Agent-computer interfaces enable automated software engineering, 2024{\natexlab{a}}.
\newblock \url{https://arxiv.org/abs/2405.15793}.

\bibitem[Yang and Zhang(2024)]{yang2024fla}
Songlin Yang and Yu~Zhang.
\newblock Fla: A triton-based library for hardware-efficient implementations of linear attention mechanism.
\newblock Software, January 2024.
\newblock \url{https://github.com/fla-org/flash-linear-attention}.

\bibitem[Yang et~al.(2024{\natexlab{b}})Yang, Kautz, and Hatamizadeh]{yang2024gateddeltanet}
Songlin Yang, Jan Kautz, and Ali Hatamizadeh.
\newblock Gated delta networks: Improving {Mamba2} with delta rule, 2024{\natexlab{b}}.
\newblock \url{https://arxiv.org/abs/2412.06464}.

\bibitem[Ye et~al.(2025)Ye, Chen, Lai, Lin, Zhang, Wang, Chen, Kasikci, Grover, Krishnamurthy, and Ceze]{ye2025flashinfer}
Zihao Ye, Lequn Chen, Ruihang Lai, Wuwei Lin, Yineng Zhang, Stephanie Wang, Tianqi Chen, Baris Kasikci, Vinod Grover, Arvind Krishnamurthy, and Luis Ceze.
\newblock Flashinfer: Efficient and customizable attention engine for llm inference serving, 2025.
\newblock \url{https://arxiv.org/abs/2501.01005}.

\bibitem[Yi et~al.(2023)Yi, Guo, Wei, Zhou, Wang, and Xu]{yi2023edgemoe}
Rongjie Yi, Liwei Guo, Shiyun Wei, Ao~Zhou, Shangguang Wang, and Mengwei Xu.
\newblock Edgemoe: Fast on-device inference of moe-based large language models, 2023.
\newblock \url{https://arxiv.org/abs/2308.14352}.

\bibitem[{Z.ai}(2026)]{glm522026}
{Z.ai}.
\newblock {GLM-5.2}.
\newblock Model checkpoint, \url{https://huggingface.co/zai-org/GLM-5.2}, 2026.

\bibitem[Zheng et~al.(2024)Zheng, Yin, Xie, Sun, Huang, Yu, Cao, Kozyrakis, Stoica, Gonzalez, Barrett, and Sheng]{zheng2024sglang}
Lianmin Zheng, Liangsheng Yin, Zhiqiang Xie, Chuyue Sun, Jeff Huang, Cody~Hao Yu, Shiyi Cao, Christos Kozyrakis, Ion Stoica, Joseph~E. Gonzalez, Clark Barrett, and Ying Sheng.
\newblock Sglang: Efficient execution of structured language model programs.
\newblock In \emph{Advances in Neural Information Processing Systems}, volume~37, pages 62557--62583, 2024.
\newblock \url{https://arxiv.org/abs/2312.07104}.

\bibitem[Zhong et~al.(2025)]{zhong2025hybrimoe}
Shuzhang Zhong et~al.
\newblock Hybrimoe: Hybrid cpu-gpu scheduling and cache management for efficient moe inference, 2025.
\newblock \url{https://arxiv.org/abs/2504.05897}.

\bibitem[Zhu et~al.(2025)Zhu, Li, Dai, Liu, Wang, Li, Xiao, Chen, and Wang]{zhu2025smoe}
Guoying Zhu, Meng Li, Haipeng Dai, Xuechen Liu, Weijun Wang, Keran Li, Jun Xiao, Ligeng Chen, and Wei Wang.
\newblock Smoe: An algorithm-system co-design for pushing moe to the edge via expert substitution, 2025.
\newblock \url{https://arxiv.org/abs/2508.18983}.

\bibitem[Zhu et~al.(2026)Zhu, Jacob, Ma, Pan, Wang, Krishnamurthy, and Kasikci]{zhu2026tracelab}
Kan Zhu, Mathew Jacob, Chenxi Ma, Yi~Pan, Stephanie Wang, Arvind Krishnamurthy, and Baris Kasikci.
\newblock Tracelab: Characterizing coding agent workloads for llm serving, 2026.
\newblock \url{https://arxiv.org/abs/2606.30560}.

\end{thebibliography}



\end{document}